\documentclass[conference]{IEEEtran}
\IEEEoverridecommandlockouts
\usepackage{cite}
\usepackage{amsmath,amssymb,amsfonts}
\usepackage{algorithmic}
\usepackage{graphicx}
\usepackage{textcomp}
\usepackage{xcolor}
\usepackage{comment}
\usepackage{tikz, pgfplots}
\usepackage{bm}
\pgfplotsset{compat=1.18}
\def\BibTeX{{\rm B\kern-.05em{\sc i\kern-.025em b}\kern-.08em
    T\kern-.1667em\lower.7ex\hbox{E}\kern-.125emX}}

\makeatletter
\newcommand{\linebreakand}{%
  \end{@IEEEauthorhalign}
  \hfill\mbox{}\par
  \mbox{}\hfill\begin{@IEEEauthorhalign}
}
\makeatother

\title{Two-Level Decorrelated Coded Modulation on the $D_4$ Lattice\\

\thanks{This research was supported in part by the U.S. Department of Commerce’s National Telecommunications and Information Administration (NTIA) under the Public Wireless Supply Chain Innovation Fund Grant Program (Award 24-60-IF2415: ASPEN - Advanced Signal Processing Enhancement for Next-Generation Open Radio Units), administered by the National Institute of Standards and Technology.}
}

\author{
\IEEEauthorblockN{Leopold Bertholet, Chloe Makdad, Stephen Mackes, Daniel Chew, Matthew Robinson}
\IEEEauthorblockA{\textit{Rampart Communications}\\
Linthicum Heights, MD, USA \\
\{lbertholet, cmakdad, stephen, dchew, matt\}@rampartcommunications.com}
}

\begin{document}

\maketitle

\begin{abstract}
We propose \textit{two-level decorrelated coding} (TLDC), a novel coded modulation scheme for the $\bm{D_4}$ lattice that combines Voronoi shaping with a two-stage decoding process to achieve lattice shaping and coding gains at low complexity. In TLDC, the decoded values of the first level allow the several random variables in the second level to become approximately uncorrelated. The resulting independence of the variables in level two permits decoding in parallel or consolidation into a larger codeword, enhancing performance. TLDC supports flexible choice of FEC within each level. Using bit-interleaved or multi-level polar codes at each level, the resulting coded modulation scheme exhibits a gain of up to 0.5 dB over analogous state-of-the-art coded modulation schemes on a 16-QAM under AWGN at block sizes of 64 and 1024 bits.
\end{abstract}

\section{Introduction}

\IEEEPARstart{T}{raditional} modulation formats such as Quadrature-Amplitude Modulation (QAM), when combined with modern binary Forward Error Correction (FEC) schemes, are sufficient to achieve capacity on AWGN channels up to shaping \cite{wachsmannMLC}. Shaping gain can be achieved by either probabilistic \cite{Kschischang1993OptimalNonuniformSignaling} or geometric \cite{mirani2020low} shaping. Voronoi constellations in the fundamental cell of a dense lattice are the most popular choice for realizing geometric shaping gains for their balance of low complexity with high performance \cite{li2025coded}, \cite{li2021designing}. In addition to shaping gain, dense lattice constellations provide coding gain, which is present without the use of FEC and provides a natural benefit over QAMs at high code rates \cite{li2021designing}, \cite{stern2019two}. Shaping with a hypersphere also exhibits gains, but is much higher complexity than Voronoi shaping \cite{frey2019coded}. What remains to be seen is a design that simultaneously modulates signal points on a lattice denser than the integer lattice, shapes by its Voronoi cell, and achieves competitive FEC gains while operating at low complexity. We accomplish this while also proposing a novel geometric design principle for multi-level codes. 

We introduce \textit{two-level decorrelated coding} (TLDC), a coded modulation for the $D_4$ lattice that exploits the existence of an orthogonal rank-three sublattice generated by short vectors. TLDC allocates two levels within each symbol. The first level is decoded as a PAM (pulse amplitude modulation), and the second is decoded as three distinct PAM symbols,  which are assumed to be uncorrelated. On each of these levels, we can place a bit-interleaved coded modulation (BICM), multi-level code (MLC), or non-binary coded modulation scheme with any underlying FEC. We opt to simulate BICM and MLC with polar codes as the underlying FEC.

\section{Modulation} \label{sec:mod}

There are two key components to TLDC modulation: a constellation and a two-level structure. We first describe the known lattice constellation and use it to define a modulation map before describing its relevant properties. We then construct the two-level structure for the data modulated into the constellation.

% We describe a known lattice constellation and then construct a two-level structure for the data modulated into it.

\subsection{Constellation}
For our constellation, we use a lattice-valued and Voronoi shaped modulation map as described in \cite{conway1983fast} on the $D_4$ lattice. One can compile bit tuples into $\mathbb{Z}/r\mathbb{Z}$ integers, hereafter called \textit{rits}, where $r$ is called the rit \textit{modulus} and is always a power of two. Four rits can be assembled to form a $D_4$ symbol as follows. Subsequently, the modulation is a map 
\begin{align*}
    \Phi : \left( \mathbb{Z}/r\mathbb{Z} \right)^4 \to D_4 \subset \mathbb{R}^4.
\end{align*}
We define the constellation $\mathcal{C}$ to be the image of $\Phi$, which has three critically relevant properties. First, $\Phi$ is surjective on the set of $D_4$ points that lie inside of the Voronoi cell $\mathcal{V}$ of $r D_4$. This property is enforced by first mapping data into $D_4$ without restricting to the Voronoi cell and then applying a quotient map $w:\mathbb{R}^4 \to \mathcal{V}$. We call the effect of this quotient of $\mathbb{R}^4$ \textit{wrapping}. The wrapping map is often referred to as a Voronoi quotient, as it uses the lattice quantizer to subtract the nearest element in a scaled sublattice. In general, applying a map $w$ to a line $\ell = \{t \bar{v}|t\in\mathbb{R}\}$ does not map $\ell$ into itself; instead, $w$ maps $\ell$ into several disconnected line segments. Second, the map $\tilde{\Phi}: \left(\mathbb{Z}/r\mathbb{Z}\right)^4 \to D_4/rD_4$, obtained by composing $\Phi$ with the quotient map $q: D_4 \to D_4/rD_4$, is a homomorphism of groups. Third, the chosen generator matrix for $D_4$ determines a map $M: \mathbb{Z}^4 \to D_4$ by matrix multiplication, and this completely determines $\tilde{\Phi}$. We use the following $D_4$ generator matrix $M$, 
\begin{align}
    M = \left( \begin{array}{c c c c}
        1 & 1 & 0 & 0 \\
        1 & -1 & 1 & 0 \\
        0 & 0 & -1 & 1 \\
        0 & 0 & 0 & -1
    \end{array} \right)
    \label{eq:d4_basis}
\end{align}
where the columns are the generating vectors.

\subsection{Two-Level Structure}\label{sec:two-level}

% We describe the construction of two levels within the constellation for TLDC. 

As the name suggests, TLDC contains two levels, which are protected by distinct codes and decoded in two stages with multistage decoding (MSD). On $D_4$, the first level consists of the third rit, which modulates into the span of the third column of $M$, while the second level comprises the remaining three rits, which are the coefficients of columns 1, 2, and 4 of $M$. This collection forms a pairwise orthogonal set, as illustrated by the Gram matrix $A = M^TM$, whose entries $A_{ij}$ are the dot products of basis vectors $i$ and $j$ \cite{conway1999sphere}:
\[
    A = \begin{pmatrix}
        2 & 0 & 1 & 0 \\
        0 & 2 & -1 & 0 \\
        1 & -1 & 2 & -1 \\
        0 & 0 & -1 & 2
    \end{pmatrix}.
\]
In particular, the sublattice of $D_4$ generated by the second level is congruent by a Euclidean isometry to $\sqrt{2} \mathbb{Z}^3$. 

Since the first, second, and fourth columns of $M$ are mutually orthogonal, the modulation first maps the three rits into a rank-three orthogonal lattice, making the rit values independent as random variables. Then, the wrapping map $w$ sends the three rits into a rank-four sublattice consisting of the original rank-three sublattice and the shifts of it by $r\mathbb{Z}$ multiples of the third column vector of $M$. We call these shifts of the three dimensional sublattice \textit{sheets}. As a result of this wrapping, the three rits are no longer independent as random variables under AWGN. However, the Euclidean distance between points in two distinct sheets is at least $\frac{r}{2}$ times greater than the minimum distance between two points in one sheet. Hence for $r > 2$ and at high SNR, the probability of points in the sheet nearest to a received point dominates that of points in other sheets. See Section \ref{sec:decor} for a numerical demonstration of this phenomenon.

There are two main design principles for multi-level schemes in the literature. Most famously, Ungerboeck's set partitioning design aims to maximally increase the minimum distance of each level \cite{ungerboeck1}. On PAMs and QAMs, the minimum distance in level $i+1$ is twice the minimum distance in level $i$. Alternatively, some designs make use of a Gray code on the overall constellation, producing levels that are similar in strength so that an error in one level is less harmful to the other levels, as is done in parallel-decoded MLC \cite{wachsmannMLC}. More recently, \cite{stern2019two} proposes a two-stage scheme on $D_4$ in which the first level is one bit determining whether two I/Q points are both on a QAM shifted `up' or a QAM shifted  `down'\footnote{The interested reader is invited to notice that the constellations from this work are a different cardinality than ours and are not Voronoi shaped, resulting in a distinct design principle.}. 

Our two-level structure employs a new design principle for choosing levels of lattice constellations, utilizing orthogonal sublattices which can be effectively considered independent. Given a $d$-dimensional lattice with generating set $\bar{b}_1, \ldots, \bar{b}_d$, we refer to a subset $S=\{\bar{b}_i\}_{i \in I}$ for some $I \subset \{1,\ldots,d\}$ as a \textit{decorrelating set} if all the vectors in $S$ are orthogonal. Decorrelating sets may be identified for $D_4$ with the basis given by the columns of $M$ in \eqref{eq:d4_basis}, as well as for other lattices. We choose the largest available decorrelating set, called a \textit{maximal} decorrelating set. 

While our focus is the $D_4$ lattice, it is worth noting that decorrelating sets also exist for other lattices. There is little to say about the $A_2$ lattice, for which any decorrelating set contains only one vector. Meanwhile, the $E_8$ lattice has decorrelating sets of order four. In fact, a basis for $E_8$ may be formed from the union of two distinct maximal decorrelating sets.

\section{$D_4$ Demodulation}

We present two classes of methods for demodulation: \textit{standard} and \textit{wrapped}. By demodulation, we mean a map from an AWGN-corrupted lattice symbol to relevant soft information, such as log-likelihood ratios or probability mass functions (PMF), which are our focus in this work. Denote the AWGN channel variance by $\sigma^2$. Standard demodulation provides optimal performance without reducing the exponential complexity of lattice modulated data, while wrapped demodulation sacrifices some coding gains in order to achieve low complexity. In particular, the complexity of a standard PMF is captured by the fact that there are $r^4$ possible states for each constellation point. Hence, the standard demodulation step computes $r^4$ real numbers, which are then marginalized to obtain PMFs of length $r$ for each of the dimensions. In contrast, the wrapped approach directly obtains PMFs for each dimension without passing through a marginalizing step, computing $4 r$ numbers. We show in Section \ref{sec:wrapped_pmfs} that each one can be calculated as the product of four real numbers evaluated from the transcendental \textit{Jacobi theta function}, which may be stored in in a look-up table (LUT). Storing that function efficiently allows the PMFs to be computed with only addition and multiplication. Loosely, we pass from $r^4$ uses of a LUT in the standard case to $16 r$ uses which compute $4 r$ numbers for wrapped.

Under MSD, the first stage of decoding demodulates received noisy data to a PMF on the first-level rit with the second level marginalized. In the second stage, PMFs on the second-level rits are conditioned on the known values in the first level and marginalize each other. This marginalizing within the second stage is not standard under MSD and is enabled by the decorrelating property of the first stage.

\subsection{Standard Gaussian Joint PMF}

The optimal PMF on the $D_4/rD_4$ constellation is computed by evaluating the Gaussian noise PDF centered on the received value at each constellation point. Let $\bar{\mu}$ denote the received value in $\mathbb{R}^4$ out of the channel. The standard PMF for all $\bar{v} \in (\mathbb{Z}/r\mathbb{Z})^4$ is
\begin{align*}
    P_{\bar{\mu}}(\bar{v}) = c \exp{\left(-\frac{1}{2\sigma^2} \| \bar{\mu} - \Phi(\bar{v}) \|^2\right)},
\end{align*}
where $c$ is the normalization constant for $P_\mu$ to be a PMF on $(\mathbb{Z}/r\mathbb{Z})^4$. The PMF on levels is then computed by appropriate marginalizing and conditioning as described above.

\subsection{Wrapped Rit PMFs}\label{sec:wrapped_pmfs}

Instead of computing the full joint PMF, which consists of $r^4$ real values, we can approximate the noise distribution by a wrapped normal distribution, which is wrapped by $rD_4$. This wrapped normal distribution can be written as a theta function in two variables on $r D_4$. Then, the relevant marginalized PMFs also arise as theta functions on various sublattices of $D_4$. These are efficiently approximated using the Jacobi theta function, which can be stored as a LUT for efficient implementation. 

The two variable theta function of a lattice $\Lambda \subset \mathbb{R}^d$ is defined for all $\bar{z} \in \mathbb{C}^d$ and $\tau \in \mathbb{C}$ such that $\Im(\tau) >0$ by 
\begin{align*}
\vartheta_\Lambda \left(\bar{z}, \tau \right) = \sum_{\bar{\lambda}^* \in \Lambda^*} \exp\left( \pi i \tau \| \bar{\lambda}^* \|^2 + 2 \pi i \ \bar{\lambda}^* \cdot  \bar{z} \right)
\end{align*}
where $\Lambda^*$ is the dual lattice of $\Lambda$. This is often called the theta function of $\Lambda^*$. It is immediate that $\vartheta_\Lambda(\bar{z} + \bar{\lambda}, \tau) = \vartheta_\Lambda(\bar{z}, \tau)$ for all $\bar{\lambda} \in \Lambda$. Furthermore, $\vartheta_\Lambda$ is well defined and analytic in both variables in the domains specified, and is uniformly bounded on $\mathbb{R}^d \subset \mathbb{C}^d$.

Now, the first level PMF can be approximated as a PMF on $D_4/\Lambda(r,3)$, where $\Lambda(r,3)$ is defined to be the lattice generated by
\begin{align*}
    M_{r,3}= \left( \begin{array}{cccc}
        1 & 1 & 0 & 0\\
        1 & -1 & r & 0\\
        0 & 0 & -r & 1\\
        0 & 0 & 0 & -1
    \end{array} \right)
\end{align*}
and $M_{r,3}$ is $M$ with the third column scaled by $r$. Notice that the cardinality of $D_4/\Lambda(r,3)$ is $r$ and that the $r$ cosets correspond to the integer coordinate of the third column of $M$ modulo $r$.

Similarly, the three PMFs describing the three rits in level two are approximated as PMFs on $D_4/\Lambda(r,\{3,j\})$ where $\Lambda(r,\{3,j\})$ is the lattice generated by $M_{r,\{3,j\}}$, the matrix $M$ with both column 3 and column $j$ scaled by $r$. The cardinality of $D_4/\Lambda(r,\{3,j\})$ is $r^2$, but in demodulating level two, the decoded value of level one is assumed. Therefore, knowing the value of the third integer coordinate in the $M$ basis restricts the values of $D_4/\Lambda(r,\{3,j\})$ to only $r$ possible values corresponding to the $j^{\text{th}}$ coordinate in the $M$ basis modulo $r$.

Define $\Phi_j : \mathbb{Z}/r\mathbb{Z} \to D_4/ rD_4$ to be $\Phi_j(n) = n Me_j \text{ mod } rD_4$, where $e_j$ is the standard basis vector of all $0$ components except for a $1$ in the $j^{\text{th}}$ component. We can define the first level wrapped demodulation by 
\begin{align*}
p_3(n | \bar{z}) =\;& 
    \vartheta \left( \frac{z_1 + z_2 - n}{2}, \frac{\tau }{2} \right)
    \vartheta \left( \frac{z_1 - z_2 + n}{2}, \frac{\tau }{2} \right) \notag \\
    &\vartheta \left( \frac{z_3 + z_4 + n}{r}, \frac{2 \tau }{r^2} \right) 
    \vartheta \left( \frac{z_3 - z_4 + n}{2}, \frac{\tau }{2} \right), 
\end{align*}
for $n \in \mathbb{Z}/r\mathbb{Z}$ and where $\tau = 2\pi i\sigma^2$. Then, assuming that the value of the third rit (level one) is $h$, we set $\tilde{\mu} = \bar{\mu}-\Phi_3(h)$. In the second level, for $j = 1,2,\text{ or }4$, we define the second level wrapped demodulations by
\begin{align*}
    p_j(n) = \vartheta_{\Lambda(r,\{3,j\})} \left( \tilde{\mu}-\Phi_j(n) , 2\pi i \sigma^2 \right).
\end{align*}
Thus, to determine $p_j(n)$ we need only to compute $\vartheta_{\Lambda(r,\{3,j\})}$ for $j=1,2,4$. These theta functions determined by $\Lambda(r,3)$ and $\Lambda(r,\{3,j\})$ can be computed in terms of the Jacobi theta function
\begin{align*}
    \vartheta (z, \tau) := \sum_{k =-\infty}^\infty \exp \left( \pi i k^2 \tau + 2 \pi i k z \right).
\end{align*}

The result is
\begin{align*}
\vartheta_{\Lambda(r,3)}(\bar{z},\tau) =& 
    \vartheta \left( \frac{z_1 + z_2}{2}, \frac{\tau }{2} \right)
    \vartheta \left( \frac{z_1 - z_2}{2}, \frac{\tau }{2} \right) \notag \\
    &\vartheta \left( \frac{z_3 + z_4}{r}, \frac{2 \tau }{r^2} \right) 
    \vartheta \left( \frac{z_3 - z_4}{2}, \frac{\tau }{2} \right),
\end{align*}
\begin{align*}
\vartheta_{\Lambda(r,\{3,1\})}(\bar{z},\tau) =& 
    \vartheta \left( \frac{z_1 + z_2+ z_3 + z_4}{2r}, \frac{\tau }{r^2} \right) \notag \\
    & \vartheta \left( \frac{z_1 + z_2 - z_3 - z_4}{2r}, \frac{\tau}{r^2} \right)  \notag \\
    & \vartheta \left( \frac{z_1 - z_2}{2}, \frac{\tau }{2} \right)  \vartheta \left( \frac{z_3 + z_4}{2}, \frac{\tau}{2} \right),
\end{align*}
\begin{align*}
\vartheta_{\Lambda(r,\{3,2\})}(\bar{z},\tau) =& 
    \vartheta \left( \frac{z_1 - z_2 - z_3 - z_4}{2r}, \frac{\tau}{r^2} \right) \notag \\
    &\vartheta \left( \frac{z_1 - z_2 + z_3 + z_4}{2r}, \frac{\tau }{r^2} \right)  \notag\\
    &\vartheta \left( \frac{z_1 + z_2}{2}, \frac{\tau }{2} \right)   \vartheta \left( \frac{z_3 - z_4}{2}, \frac{\tau}{2} \right),
\end{align*}
\begin{align*}
\vartheta_{\Lambda(r,\{3,4\})}(\bar{z},\tau) =& 
    \vartheta \left( \frac{z_1 + z_2}{2}, \frac{\tau }{2} \right)
    \vartheta \left( \frac{z_1 - z_2}{2}, \frac{\tau}{2} \right) \notag \\
    &\vartheta \left( \frac{z_3}{r}, \frac{\tau}{r^2} \right) 
    \vartheta \left( \frac{z_4}{r}, \frac{\tau}{r^2} \right). \\
\end{align*}
An efficiently implemented (e.g., LUT-based) Jacobi theta function makes these much less computationally intensive than the standard Gaussian PMFs.

\section{Decorrelation Demonstrated}\label{sec:decor}

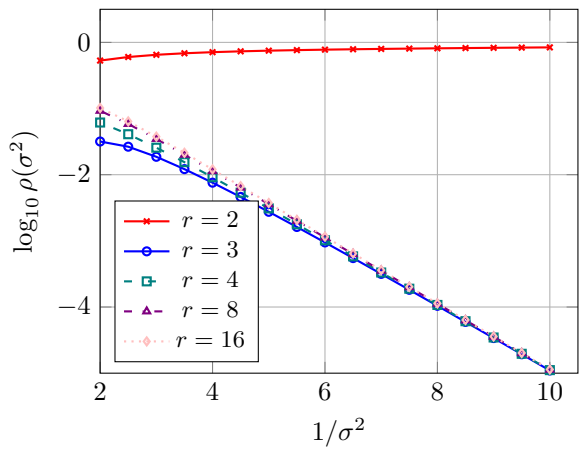
\begin{figure}
    \centering
    \begin{tikzpicture}
\begin{axis}[
    scale = 0.75,
    name = decorrelation,
    xshift=2cm,
    xlabel={$1/\sigma^2$},
    ylabel={$\log_{10}\rho(\sigma^2)$},
    xmin=2, xmax=10.5,
    ymin=-5, ymax=0.5,
    grid=both,
    grid style={line width=0.2pt, draw=gray!30},
    major grid style={line width=0.4pt, draw=gray!60},
    width=10cm,
    height=8cm,
    mark size=1.5pt,
    legend style={font=\small, at={(0.03,0.03)},anchor= south west}
]

\addplot[
    color=red,
    mark=x,
    mark options = solid,
    solid,
    thick,
] coordinates {
    (1., -0.630137)
    (1.5, -0.376557)
    (2., -0.274899)
    (2.5, -0.221129) 
    (3., -0.187603)
    (3.5, -0.164648) 
    (4., -0.147949) 
    (4.5, -0.135229)
    (5., -0.125159) 
    (5.5, -0.11692) 
    (6., -0.109986) 
    (6.5, -0.104007) 
    (7., -0.0987469) 
    (7.5, -0.0940385)
    (8., -0.089763)
    (8.5, -0.0858342)
    (9., -0.0821883) 
    (9.5, -0.0787774) 
    (10., -0.0755657) 
};
\addlegendentry{$r=2$}

\addplot[
    color=blue,
    mark=o,
    mark options = solid,
    solid,
    thick,
] coordinates {
    (1., -1.38668)
    (1.5, -1.62513)
    (2., -1.49823)
    (2.5, -1.57672)
    (3., -1.72938)
    (3.5, -1.91573)
    (4., -2.12052)
    (4.5, -2.33679)
    (5., -2.56079)
    (5.5, -2.79028)
    (6., -3.02381)
    (6.5, -3.26037)
    (7., -3.49925)
    (7.5, -3.73992)
    (8., -3.98199)
    (8.5, -4.22516)
    (9., -4.4692)
    (9.5, -4.71392)
    (10., -4.9592)

};
\addlegendentry{$r=3$}

\addplot[
    color=teal,
    mark=square,
    mark options = solid,
    loosely dashed,
    thick,
] coordinates {
    (1., -1.11644)
    (1.5, -1.08825)
    (2., -1.21142)
    (2.5, -1.38815)
    (3., -1.5917)
    (3.5, -1.81041)
    (4., -2.03826)
    (4.5, -2.27189 )
    (5., -2.50931)
    (5.5, -2.74932)
    (6., -2.99114)
    (6.5, -3.23428)
    (7., -3.47839)
    (7.5, -3.72322)
    (8., -3.96862)
    (8.5, -4.21444)
    (9., -4.4606)
    (9.5, -4.70706)
    (10., -4.95378)

};
\addlegendentry{$r=4$}

\addplot[
    color=violet,
    mark=triangle,
    mark options = solid,
    loosely dashdotted,
    thick,
] coordinates {
    (1., -0.868101)
    (1.5, -0.890207)
    (2., -1.03352)
    (2.5, -1.23259)
    (3., -1.4587)
    (3.5, -1.69839)
    (4., -1.9449)
    (4.5, -2.19468)
    (5., -2.44585)
    (5.5, -2.69741)
    (6., -2.94883)
    (6.5, -3.1999)
    (7., -3.45051)
    (7.5, -3.70066)
    (8., -3.95038)
    (8.5, -4.19972)
    (9., -4.44874)
    (9.5, -4.69751)
    (10., -4.94605)

};
\addlegendentry{$r=8$}

\addplot[
    color=pink,
    mark=diamond,
    mark options = solid,
    dotted,
    thick,
] coordinates {
    (1., -0.825782)
    (1.5, -0.846334)
    (2., -0.991519)
    (2.5, -1.19441)
    (3., -1.42498)
    (3.5, -1.66914)
    (4., -1.91988)
    (4.5, -2.17352)
    (5., -2.42812)
    (5.5, -2.68266)
    (6., -2.93665)
    (6.5, -3.18988)
    (7., -3.44231)
    (7.5, -3.69397)
    (8., -3.94494)
    (8.5, -4.19533)
    (9., -4.4452)
    (9.5, -4.69458)
    (10., -4.94365)
    % (10.5, -5.1939)
    % (11., -5.4806)
    % (11.5, -5.83704)
    % (12., -6.53062)
    % (12.5, -6.7667)
    % (13., -5.00497)
};
\addlegendentry{$r=16$}

\end{axis}

\end{tikzpicture}
    \caption{Plots of $\log_{10} \rho(\sigma^2)$ for second-level rits as a function of $1/\sigma^2$ for several $r$ values.}
    \label{fig:decor}
\end{figure}

In Section \ref{sec:two-level}, we showed that decorrelation occurs for $r>2$. Figure \ref{fig:decor} illustrates the rate of decay of correlation in SNR for various values of $r$. We describe the dependence of decorrelation on $r$ and SNR by computing the average Pearson correlation,  $\rho(\sigma^2)$, between two of the rits in level two for a fixed noisy received point, where $\sigma^2$ is the noise variance. For several $r$ values, we plot the $\log_{10} \rho (\sigma^2)$ for rits one and two as a function of $1/\sigma^2$. The curves do not depend on which two rits are chosen. We use wrapped PMFs in this simulation because the wrapping is the source of failure of full independence.

The curves in Figure \ref{fig:decor} show that the correlation coefficient decays exponentially at a rate independent of $r$ in SNR at all $r>2$, and that it does not decay for $r=2$. We include data for $r=3$ even though it is not immediately useful for bit transmission, as it reveals that $r>2$ is the tightest bound available for achieving decorrelation. While it appears that $r=3$ is the fastest decorrelating, this is only due to the lower variance of the higher $r$ PMFs and is counterbalanced by the smaller domain on which the decorrelation is so strong. Hence, at larger $r$, the decorrelation is stronger in the sense that more of the receivable points in $\mathbb{R}^n$ allow for level two to be effectively decorrelated. We note that this was calculated for one randomly chosen point near the origin. When the received point is farther away from its nearest sheet, the decorrelation decays more slowly.

\section{Code Design}

The two key decisions made in the code design for TLDC decoding are the choice of a code rate in each level and which component coded modulation to use. In particular, for $r>2$, multiple bits are modulated into each rit. Thus, each level contains the data of a coded modulation rather than simply a code. Any of the three standard classes of coded modulation (bit-interleaved, multi-level, or non-binary) may be used in each level. In this design, one should view the first level as an $r$-PAM and the second level as three independent $r$-PAMs. This naturally allocates levels for MLC components. In Section \ref{sec:results}, we specifically simulate with BICM polar codes in each level and with MLC polar codes in each level. However, TLDC is also compatible with any other PAM coded modulation.

Polar codewords are naturally $2^q$ bits for some integer $q$. We chose $r=4$ and use $N$-bit codewords, so there will be $N/8$ lattice symbols involved. This means that $N/4$ bits will be in each level one codeword and $3N/4$ bits will be in each level two codeword. Rather than puncturing or shortening polar codes to achieve this multiple of three block size, we use a $3\times 3$ kernel to combine the three rits of level two and achieve a $3\cdot 2^q$ bit block size polar code \cite{gabry2017multi}, \cite{saha2020versatile}. This is needed for both MLC and BICM, as the factor of three is due only to the cardinality of level two.

Code construction on multi-level codes may be achieved by allocating level rates based on the capacity of each code level \cite{wachsmannMLC}. However, the use of polar codes in conjunction with multi-level coding enables polar reliability sequence design techniques to be applied to the entire scheme \cite{seidl2013polar}. We apply these methods to  construct a reliability sequence ordering the strength of all bits in the codeword on both levels. Then, the selection of a rate provides a frozen set for each level, which implies a rate for each.

The code construction for BICM is usually done without the knowledge of the modulation, so the only choice to make is the code rates assigned to each level. We use a numerical search similar to the multi-level code construction to accomplish this.

\section{Experiments and Results}\label{sec:results}

\begin{figure*}[htbp]
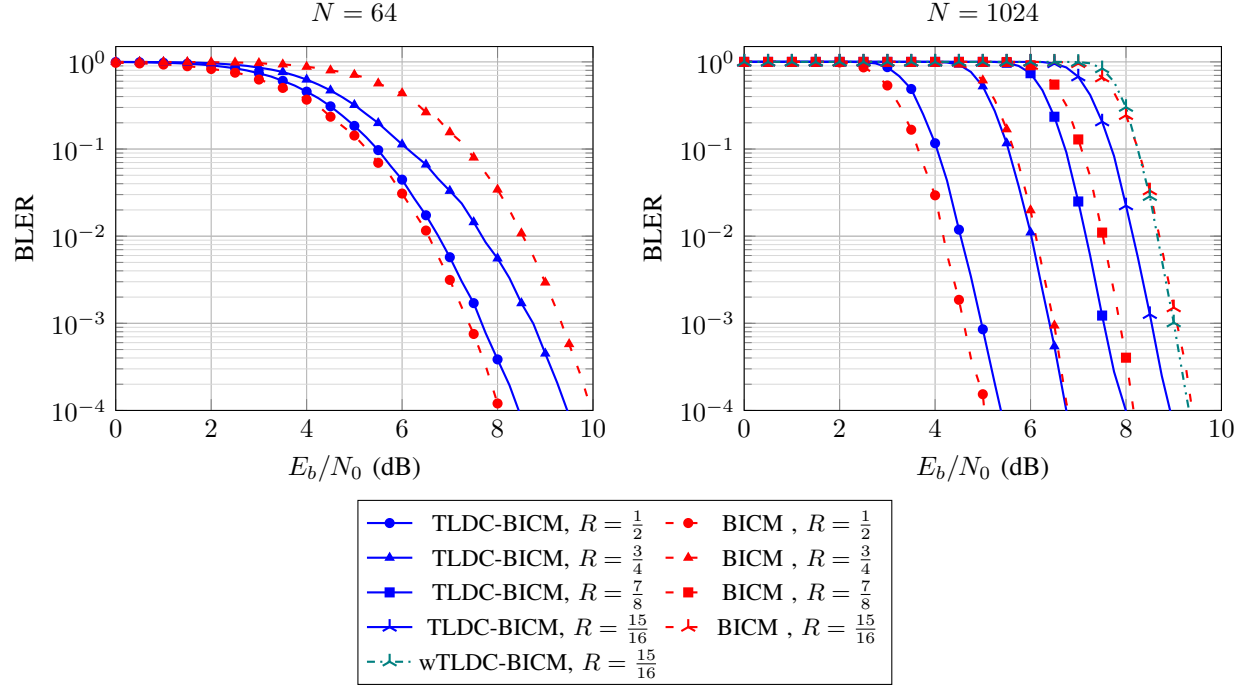

    \centering
    \include{bler_bicm}
    \caption{A comparison of TLDC-BICM with bit-interleaved polar coded modulation on a 16-QAM (BICM) in BLER at bit block sizes $N=64$ (left) and $N=1024$ (right).}
    \label{fig:blerbicm}
\end{figure*}

\begin{figure*}[htbp]
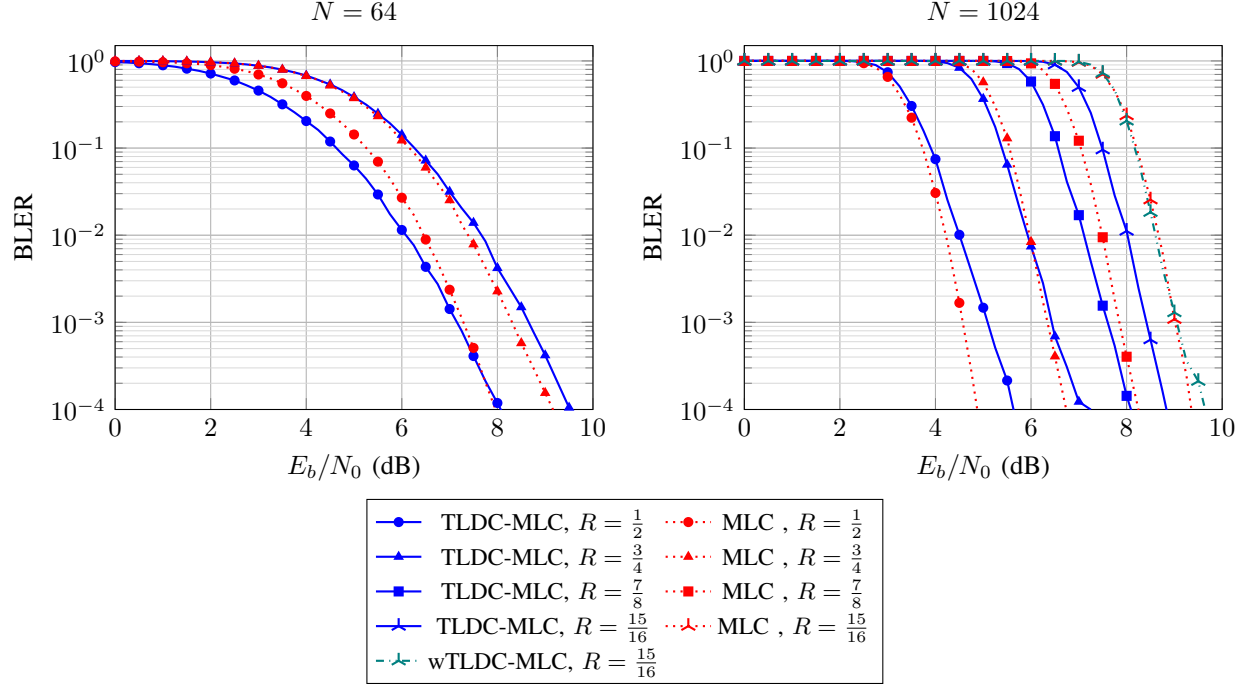

    \centering
    \include{bler_mlc}
    \caption{A comparison of TLDC-MLC with multi-level polar coded modulation on a 16-QAM (MLC) in BLER at bit block sizes $N=64$ (left) and $N=1024$ (right).}
    \label{fig:blermlc}
\end{figure*}

We compare TLDC on the $D_4$ lattice to two state-of-the-art coded modulations: MLC and BICM. Recently, BICM \cite{ciareBICM} has become the industry standard due to its ease of implementation with binary FEC and its resilience to all channel types. As such, we consider BICM to be the benchmark for comparing new coded modulation schemes. When designing only for AWGN, it is known that MLC is the state of the art, outperforming BICM \cite{wachsmannMLC}. When comparing TLDC with BICM, we use BICM on each level, and denote this as TLDC-BICM, with results in Figure \ref{fig:blerbicm}. Similarly, when comparing with MLC, we use MLC on each level and denote this as TLDC-MLC, as seen in Figure \ref{fig:blermlc}. All demodulations will be standard unless marked with a `w' for wrapped (e.g. wTLDC-BICM).

Explicitly, we evaluate each coded modulation with polar codes as the underlying FEC. All codewords will use $N$ bits. For our $D_4$ TLDC schemes, we use a polar coded modulation with rit block size $N/4$ on level one, and then a polar coded modulation with rit block size $3N/4$ on level two. The results are shown by for block error rate (BLER) simulations at bit block sizes $N = 64$ and $N = 1024$. Here, a block refers to all $N$ bits. Each decoder uses CRC-aided successive cancellation list decoding (CA-SCL) with a list size of 8. List branch data and scores are passed across decoding levels as proposed in \cite{zhang2021path}. Our $D_4$ TLDC is for modulus $r=4$. To ensure a fair comparison in terms of spectral efficiency, the BICM and MLC results are for a 16-QAM.

The results for the BICM experiments are shown in Figure \ref{fig:blerbicm}. At block size $N=64$, TLDC-BICM outperforms BICM at rate $3/4$ but underperforms at rate $1/2$. At block size $N=1024$, TLDC-BICM outperforms BICM at rate $3/4$ and higher, and underperforms at rates $1/2$ and lower. We use a wrapped PMF demodulation at rate $15/16$ to show that wTLDC-BICM is competitive with BICM in at least the very high rate schemes at large block size. In any of the other schemes, wTLDC-BICM does not outperform its competition on 16-QAM (e.g. the wTLDC-BICM curve for block size 64).

Figure \ref{fig:blermlc} shows the results for the MLC experiments. At small block size, the tradeoff in performance between MLC and TLDC-MLC varies with code rate. At block size $N=1024$, TLDC-MLC outperforms MLC at rates greater than $7/8$, but underperforms at rates less than $3/4$. In all cases except rate $15/16$, wTLDC-MLC underperforms, while at rate $15/16$, wTLDC-MLC matches the performance of MLC. In its strongest configurations exhibited here, TLDC shows gains of approximately 0.5 dB over MLC and BICM.

We highlight that TLDC generally achieves coding gains over BICM and MLC at rates greater than $3/4$. This is explained by the fact that while shaping gain is present at all rates, the lattice coding gain becomes increasingly visible at higher rates.

\section{Discussion and Future Work}

TLDC uses a novel design principle for multi-level coded modulation tailored to the $D_4$ lattice and achieves the shaping gain of $D_4$ Voronoi constellations. The resulting decoder operates at low complexity up to the choice of demodulation. We have also proposed a low-complexity demodulation scheme that comes at the cost of some performance gains by making use of theta functions.

As shown in Section \ref{sec:results}, TLDC underperforms at low rate. This is likely due to the much lower rate that the capacity rule and our numerical code construction place in the first level. In particular, the knowledge of the first level does not significantly improve the certainty of the second. This contrasts with more standard MLC construction using set partitioning, where if the first level is rate 0 or very low, better performance would be driven by increased certainty on the second level. In fact, for TLDC, the knowledge of the first level only decorrelates the second, so the low rate performance is lacking at any finite block size.

Natural avenues for future work include evaluating TLDC in fading channels and with non-binary component FEC. We also hope to improve the performance of the wrapped demodulation or reduce the complexity of standard demodulation for realistic applications.

As mentioned, this strategy immediately yields a two-stage decorrelating coded modulation on $E_8$. While the two-stage model is designed for $D_4$, the orthogonality properties available in $E_8$ bases in conjunction with wrapped demodulation methods may reveal novel low-complexity coded modulations using the decorrelation design principle.

\bibliographystyle{IEEEtran}
\bibliography{bibliography}

@article{li2025coded,
  title={Coded modulation schemes for Voronoi constellations},
  author={Li, Shen and Mirani, Ali and Karlsson, Magnus and Agrell, Erik},
  journal={IEEE Transactions on Communications},
  year={2025},
  publisher={IEEE}
}

@inproceedings{stern2019two,
  title={Two-stage dimension-wise coded modulation for four-dimensional Hurwitz-integer constellations},
  author={Stern, Sebastian and Frey, Felix and Fischer, Johannes K and Fischer, Robert FH},
  booktitle={SCC 2019; 12th International ITG Conference on Systems, Communications and Coding},
  pages={1--6},
  year={2019},
  organization={VDE}
}

@inproceedings{frey2019coded,
  title={Coded modulation using a 512-ary Hurwitz-integer constellation},
  author={Frey, Felix and Stern, Sebastian and Emmerich, Robert and Schubert, Colja and Fischer, Johannes K and Fischer, Robert FH},
  booktitle={45th European Conference on Optical Communication (ECOC 2019)},
  year={2019},
  organization={IET}
}

@inproceedings{zhang2021path,
  title={Path metric inherited SCL decoding of multilevel polar-coded systems},
  author={Zhang, Dexin and Wu, Bolin and Niu, Kai},
  booktitle={2021 IEEE wireless communications and networking conference workshops (WCNCW)},
  pages={1--6},
  year={2021},
  organization={IEEE}
}

@ARTICLE{conway1983fast,
  author={Conway, J. and Sloane, N.},
  journal={IEEE Transactions on Information Theory}, 
  title={A fast encoding method for lattice codes and quantizers}, 
  year={1983},
  volume={29},
  number={6},
  pages={820-824},
  doi={10.1109/TIT.1983.1056761}}

@ARTICLE{wachsmannMLC,
  author={Wachsmann, U. and Fischer, R.F.H. and Huber, J.B.},
  journal={IEEE Transactions on Information Theory}, 
  title={Multilevel codes: theoretical concepts and practical design rules}, 
  year={1999},
  volume={45},
  number={5},
  pages={1361-1391},
  doi={10.1109/18.771140}}

@book{conway1999sphere,
  author    = {Conway, John H. and Sloane, Neil J. A.},
  title     = {Sphere Packings, Lattices and Groups},
  edition   = {3rd},
  publisher = {Springer},
  year      = {1999},
  series    = {Grundlehren der mathematischen Wissenschaften},
  volume    = {290},
  isbn      = {978-0-387-98585-5},
  doi       = {10.1007/978-1-4757-6568-7}
}

@ARTICLE{ungerboeck1,
  author={Ungerboeck, G.},
  journal={IEEE Transactions on Information Theory}, 
  title={Channel coding with multilevel/phase signals}, 
  year={1982},
  volume={28},
  number={1},
  pages={55-67},
  doi={10.1109/TIT.1982.1056454}}

@INPROCEEDINGS{ciareBICM,
  author={Caire, G. and Taricco, G. and Biglieri, E.},
  booktitle={Proceedings of IEEE International Symposium on Information Theory}, 
  title={Bit-interleaved coded modulation}, 
  year={1997},
  volume={},
  number={},
  pages={96-},
  doi={10.1109/ISIT.1997.613011}}

@inproceedings{li2021designing,
  title={Designing Voronoi constellations to minimize bit error rate},
  author={Li, Shen and Mirani, Ali and Karlsson, Magnus and Agrell, Erik},
  booktitle={2021 IEEE International Symposium on Information Theory (ISIT)},
  pages={1017--1022},
  year={2021},
  organization={IEEE}
}

@article{mirani2020low,
  title={Low-complexity geometric shaping},
  author={Mirani, Ali and Agrell, Erik and Karlsson, Magnus},
  journal={Journal of Lightwave Technology},
  volume={39},
  number={2},
  pages={363--371},
  year={2020},
  publisher={IEEE}
}

@article{seidl2013polar,
  title={Polar-coded modulation},
  author={Seidl, Mathis and Schenk, Andreas and Stierstorfer, Clemens and Huber, Johannes B},
  journal={IEEE Transactions on Communications},
  volume={61},
  number={10},
  pages={4108--4119},
  year={2013},
  publisher={IEEE}
}

@inproceedings{gabry2017multi,
  title={Multi-kernel construction of polar codes},
  author={Gabry, Fr{\'e}d{\'e}ric and Bioglio, Valerio and Land, Ingmar and Belfiore, Jean-Claude},
  booktitle={2017 IEEE International Conference on Communications Workshops (ICC Workshops)},
  pages={761--765},
  year={2017},
  organization={IEEE}
}

@inproceedings{saha2020versatile,
  title={Versatile Polar Codes with different Kernel Sizes and Rate Matching approaches},
  author={Saha, Souradip and Adrat, Marc},
  booktitle={2020 14th International Conference on Signal Processing and Communication Systems (ICSPCS)},
  pages={1--7},
  year={2020},
  organization={IEEE}
}

@ARTICLE{Kschischang1993OptimalNonuniformSignaling,
  author={Kschischang, F.R. and Pasupathy, S.},
  journal={IEEE Transactions on Information Theory}, 
  title={Optimal nonuniform signaling for Gaussian channels}, 
  year={1993},
  volume={39},
  number={3},
  pages={913-929},
  doi={10.1109/18.256499}}

\end{document}